\documentclass[twocolumn,superscriptaddress,nofootinbib]{revtex4}

\usepackage{amsmath}
\usepackage{epsfig}
\usepackage{graphics}
\usepackage{graphicx}
\usepackage[colorlinks=true, citecolor=blue, urlcolor = blue, linkcolor= red, bookmarks=true]{hyperref}

\begin{document}
\title{Shadow of charged wormholes in Einstein-Maxwell-dilaton theory}
\author{Muhammed Amir}
\email{amirctp12@gmail.com}
\affiliation{Astrophysics and Cosmology Research Unit, School of Mathematics, 
Statistics and Computer Science, University of KwaZulu-Natal, 
Private Bag X54001, Durban 4000, South Africa}

\author{Ayan Banerjee}
\email{ayan\_7575@yahoo.co.in}
\affiliation{Astrophysics and Cosmology Research Unit, School of Mathematics,
Statistics and Computer Science, University of KwaZulu-Natal,
Private Bag X54001, Durban 4000, South Africa}

\author{Sunil D. Maharaj}
\email{maharaj@ukzn.ac.za}
\affiliation{Astrophysics and Cosmology Research Unit, School of Mathematics,
Statistics and Computer Science, University of KwaZulu-Natal,
Private Bag X54001, Durban 4000, South Africa}

\date{\today }

\begin{abstract}
The study of shadow is quite prominent nowadays because of the ongoing Event Horizon 
Telescope\footnote{https://eventhorizontelescope.org/} observations. We construct the shadow images
of charged wormholes in Einstein-Maxwell-dilaton (EMD) theory. The spacetime metric of the charged wormholes 
contains three charges: magnetic charge $P$, electric charge $Q$, and dilaton charge $\Sigma$. We evaluate 
the photon geodesics around the charged wormholes. We also calculate the effective potential and discuss 
its behavior with angular momentum $L$ and different values of charges $P$, $Q$, and $\Sigma$. A study of 
the shadow of charged wormholes reveals that the shadow has an effect of the charges $P$ and $Q$. 
The radius of the shadow increases with the magnetic charge $P$ as well as 
the electric charge $Q$. We also find that the dilaton charge does not affect the shadow of the charged 
wormholes.
\end{abstract}

\maketitle

\section{Introduction}
\label{int}
Wormholes are one of the most interesting predictions of general relativity, which act as tunnel-like 
spacetime structures connecting different points in spacetime or even in different universes. To date 
wormholes are largely considered as in science fiction, though it has a long history after Einstein 
proposed the general theory of relativity. From a theoretical perspective, wormhole physics can 
originally be traced back to Flamm in 1916 \cite{Flamm}, soon after the discovery of the 
Schwarzschild solution. Further, wormhole type solutions were considered by Einstein and Rosen 
\cite{Einstein:1935tc} in 1935, they wanted to rid physics of singularities, and they introduced a 
bridge-like structure connecting two identical sheets. This mathematical construction  came to be 
called ``Einstein-Rosen bridges". Some time ago Wheeler \cite{WH,Fuller:1962zza} revived the subject 
and discussed wormholes that connected different regions of spacetime in terms of topological 
entities called geons, which were transformed later into Euclidean wormholes by Hawking 
\cite{Hawking:1988ae} and others.

Modern interest in wormholes is mainly based on the pioneering work of Morris and Thorne \cite{MM}, 
where observers may freely traverse, and the way to convert them into time machines \cite{Morris:1988tu}. They 
proposed wormholes as a tool for teaching general relativity, and for attracting young students into 
the field. These developments were boosted by the publication of the book ``Lorentzian Wormholes: 
From  Einstein to Hawking" by Visser \cite{MV}, where the author reviewed the subject up to 1995, as 
well as proposing several new ideas. To make a Lorentzian wormhole traversable, it has to be necessarily 
violate the null energy condition (NEC), at least in a neighborhood of the wormhole throat 
\cite{MM,Hochberg}, according to the needs of the geometrical structure. Regarding the energy 
conditions, the NEC is the weakest of the energy conditions, which implies the violation of other 
energy conditions also. However, such a matter appears in quantum field theory, since in the Casimir 
effect the null, weak, strong, and dominant energy conditions are all violated. 

Various efforts have been taken to avoid the use of exotic matter, but all were in vain within the 
context of general relativity. Nevertheless, a popular approach had been taken by Visser 
\emph{et al.} \cite{Dadhich1}, namely the ``volume integral quantifier" which quantifies the total 
amount of energy condition violating matter. Nandi \emph{et al.} \cite{Nandi} suggested an exact 
integral quantifier for matter violating averaged null energy condition (ANEC). During the last 
decade, the existence of sufficiently stable wormhole solutions have been found 
in support of phantom-like matter \cite{Sushkov,Lobo11,Lobo22,Rashid}. An alternative interpretation 
was given in \cite{Rahaman} with a variable equation of state parameter $\omega(r)$. On the other 
hand, dynamic wormhole geometries were analyzed in specific cases \cite{Kar,Kar1,Anchordoqui}, 
whereas self-dual Lorentzian wormholes have been studied in \cite{Cataldo,Dadhich:2001fu}. 
While searching the evidences regarding the existence of wormholes, gravitational lensing 
has been intensively studied by the astrophysical community. Gravitational lensing for wormholes in the 
strong/weak field approximation has been developed in \cite{n0,n1,n2,n3,n4,n5,n6,n7,n8,n9,n10,n11,n12,n13}.

Based on these analyses, the most challenging problem is to construct traversable and stable 
wormhole solutions within classical gravitational physics. In this regard a large number of work 
have been devoted to model and study wormhole geometries within the context of modified gravity 
including $f(R)$ gravity \cite{Lobo3}, Born-Infeld theory \cite{Eiroa}, noncommutative 
geometry \cite{Sharif}, bumblebee gravity \cite{Ovgun}, and others. Also another gravity 
theory, namely, Einstein-Maxwell-Dilaton (EMD) theory, which provides various distinctive physical 
properties depending on the parameters present, have been extensively explored in different 
literatures. The field content of EMD theories have a $U(1)$ gauge field $A_{\mu}$ and a dilaton 
$\phi$, in addition to a metric $g_{\mu\nu}$, and the dilaton couples exponentially to the field 
strength. Our interest was triggered by the work of Goulart \cite{Goulart1}, who found zero mass 
point-like solutions and charged wormholes that arise from the dyonic black hole solution in 
the development of EMD theory.  Furthermore, wormholes geometries were analysed \cite{Goulart2}, and 
the deflection of light by charged wormholes \cite{Banerjee} was studied within the same context.

It is widely believed that most galaxies contain supermassive black holes at their centers 
\cite{Rees:1984si,Kormendy:1995}, e.g. Milky Way and Messier 87 have, namely, Sgr A$^*$ and M87. 
This opens the door toward the observations because direct image of these supermassive black holes 
will give strong evidence regarding their existence. Therefore the study of black hole shadow has 
received much attention in recent years. Black hole casts a shadow on the bright background as an 
optical appearance due to the strong gravitational lensing effect. Actually, an observer images the 
photon orbits around the event horizon. The Event Horizon Telescope (EHT) is the available 
observational setup regarding the detection of black hole shadow. 
BlackHoleCam\footnote{https://blackholecam.org/} is collaboratively working with the EHT 
astronomers team with the aim to make the first ever direct image of the supermassive black hole 
Sgr A$^*$. Note that an observation of Sgr A$^*$ requires an earth-sized telescope which is 
impossible to create. This issue can be resolved with the help of Very Long Baseline Interferometry 
(VLBI) network. VLBI is an astronomical interferometry in which earth-based multiple radio 
telescopes collect signals from astronomical radio sources. These radio telescopes are distributed 
across different parts of earth such that a virtual earth-sized telescope can be constructed, 
and it is able to produce highest resolution at the mm/sub-mm wavelength scale. 
Schwarzschild black hole shadow is discussed by Synge \cite{Synge:1966} and it is further
explored by Luminet \cite{Luminet:1979}. Bardeen \cite{Bardeen:1973gb,Chandrasekhar:1992} studied the 
shadow cast by the Kerr black holes and further extension was provided by Falcke 
\cite{Falcke:1999pj}. The available literature on this topic contains the study of shadow for various 
types of black holes 
\cite{Takahashi:2005hy,Schee:2008kz,Bambi:2008jg,Hioki:2008zw,Hioki:2009na,Wei:2013kza,Bambi:2010hf,Amarilla:2010zq,Amarilla:2011fx,Amarilla:2013sj,Yumoto:2012kz,Abdujabbarov:2012bn,Atamurotov:2013dpa,Atamurotov:2013sca,Li:2013jra,Grenzebach:2014fha,Cunha:2015yba,Tinchev:2015apf,Johannsen:2015qca,Abdujabbarov:2016hnw,Amir:2016cen,Kumar:2017vuh,Singh:2017xle}. 
Naked singularity shadows are discussed in \cite{Hioki:2009na,Amarilla:2011fx,Papnoi:2014aaa,Amir:2017slq}. 
Moreover, this phenomenon has also been considered in higher-dimensional black holes 
\cite{Papnoi:2014aaa,Amir:2017slq,Singh:2017vfr}, and the authors in \cite{Abdujabbarov:2015xqa,Goddi:2016jrs,Younsi:2016azx} 
considered different approaches to explore it more precisely. 

It is noticed that apart from the black holes and the naked singularities, the wormholes also cast shadow 
\cite{Nedkova:2013msa,Bambi:2013nla,AzregAinou:2014dwa,Ohgami:2015nra,Ohgami:2016iqm,Abdujabbarov:2016efm,Shaikh:2018kfv}, 
which could be helpful to probe the physics of wormholes. Taking motivation from the recent activities on the shadow of 
wormholes, we are going to construct the shadow of charged wormholes in EMD theory, and explicitly bring out the effect 
of charges on the shapes of the shadow.

The paper is organized as follows. In Sec.~\ref{worm}, we briefly review the charged wormholes 
solution in EMD theory. Section~\ref{geodesics} is devoted to the calculation of photon geodesics 
around the charged wormholes and the evaluation of impact parameters required for the study of 
shadow. The shadow study of charged wormholes is the subject of Sec.~\ref{wshadow}. Finally, we 
conclude our results in Sec.~\ref{concl}. We have used geometrized units, $G=c=1$ throughout this 
paper.

\section{Dyonic Wormholes in the Einstein-Maxwell-dilaton theory}
\label{worm}
We start with a brief review about dyonic wormholes in the EMD theory \cite{Goulart1,Goulart2}. 
The authors considered the EMD theory without a dilaton potential and the corresponding action is 
written as
\begin{equation} \label{ad}
S = \int d^{4}x\sqrt{-g}\left(R-2\partial_{\mu}\phi\partial^{\mu}\phi
-W(\phi)F_{\mu\nu}F^{\mu\nu}\right),
\end{equation}
where $\phi$ denotes the dilaton scalar field, $R$ is the Ricci scalar, and $F_{\mu \nu}$ represents 
the electromagnetic field strength with the form
\begin{equation}
F_{\mu\nu}=\partial_{\mu}A_{\nu}-\partial_{\nu}A_{\mu},
\end{equation}
where $ A_{\nu}$ is the gauge potential. We consider the field equations for the metric, dilaton, and 
gauge fields, and Bianchi identities arising from the action (\ref{ad}) in the form
\begin{equation}
R_{\mu\nu}=2\partial_{\mu}\phi\partial_{\nu}\phi-\frac{1}{2}g_{\mu\nu} 
W(\phi)F_{\rho\sigma}F^{\rho\sigma}+2W(\phi)F_{\mu\rho}F_{\nu}^{\rho},
\end{equation}
\begin{equation}
\nabla_{\mu}\left(\partial^{\mu} \phi\right)-\frac{1}{4}\frac{\partial W(\phi)}
{\partial \phi}F_{\mu\nu}F^{\mu\nu} =0,
\end{equation}
\begin{equation}
\nabla_{\mu}\left(W(\phi)F^{\mu\nu}\right) =0,
\end{equation}
\begin{equation}
\nabla_{[\mu}{F_{\rho\sigma]}} =0.
\end{equation}
Among all solutions of the theory, we only consider here the particular case $W(\phi)=e^{-2\phi}$ 
\cite{Goulart1}. With this input, one can obtain the bosonic sector $SU(4)$ version of $N = 4$ 
supergravity theory \cite{Cremmer} for constant axion field. An argument by Kallosh \emph{et al.} 
\cite{Kallosh}, showed that  the extreme solutions saturate the supersymmetry bound of $N = 4, d = 4$ 
supergravity, or dimensionally reduced superstring theory. Moreover, the references 
\cite{Koikawa,Gibbons,Yoshimura} 
include other solutions related to black holes where it was shown that the dilaton changes the 
causal structure of the black hole which leads to the curvature singularities at finite radii. Our 
interest in the problem was greatly enhanced by the result that low energy limit of string theory  
includes a scalar dilaton field, which is massless in all finite orders of perturbation theory 
\cite{Green}. The 
underlying non-extremal dyonic black hole solution for Einstein-Maxwell-dilaton theory, in absence 
of a scalar potential, with integration constants which is constrained by the equations of motion, 
can be expressed as \cite{Goulart1}:
\begin{equation}\label{genmet}
\mathrm{d}s^{2}=-f(r)\mathrm{d}t^{2}+\frac{1}{f(r)}\mathrm{d}r^{2}+h(r)\mathrm{d}\Omega_2^2. 
\end{equation}
In the above $\mathrm{d}\Omega_2^2 = \mathrm{d}\theta^{2}+\sin^{2}\theta \mathrm{d}\varphi^{2}$ denotes the 
line element of the unit 2-sphere with the metric functions 
\begin{align}
f(r) & =\frac{(r-r_{1})(r-r_{2})}{(r+d_{0})(r+d_{1})},\,\,\,h(r)=(r+d_{0})(r+d_{1}),\label{genmets}\\
e^{2\phi} & =e^{2\phi_{0}}\frac{r+d_{1}}{r+d_{0}},\label{gendil}\\
F_{rt} & =\frac{e^{2\phi_{0}}Q}{(r+d_{0})^{2}},\,\,\,F_{\theta\phi}=P\sin\theta,\label{genele}
\end{align}
where the parameters $Q$, $P$ and $\phi_0$ are electric charge, magnetic charge, and 
the value of the dilaton at infinity, with four integrating constant $r_1$, $r_2$, $d_0$ and $d_1$. 
Note that this solution is free of boundary conditions.

We are interested here in a massless point like dyonic solution. Thus, for such solutions,  we consider the case when $d_{1}= -d_{0}= -\Sigma$ and $r_{1}= -r_{2} \equiv r_{H}$, to obtain the
following relation \cite{Goulart1} 
\begin{equation}\label{a}
e^{2\phi_{0}} =\pm \frac{P}{Q}.
\end{equation}

Following the argument in \cite{Goulart1}, we only consider the case of a negative sign, to 
achieve the massless solution obtained in EMD theory. For that purpose, the non-extremal solution 
corresponding to the negative sign of Eq. (\ref{a}), leads to 
\begin{align}
f(r) & =\frac{(r-r_{+})(r-r_{-})}{(r^{2}-\Sigma^{2})},\,\,h(r)=(r^{2}-\Sigma^{2}), \label{aa}\\
e^{2\phi} & =-\frac{P}{Q}\left(\frac{r-\Sigma}{r+\Sigma}\right),\\
F_{rt} & = -\frac{P}{(r+\Sigma)^{2}},\,\,\,F_{\theta\phi}=P\sin\theta.
\end{align}
The horizon and singularity are located at
\begin{equation}
r_{+}= +\sqrt{\Sigma^2+2QP}, ~~~r_{S} = |\Sigma|.
\end{equation}
Note that this solution excludes $r_{-}= -\sqrt{\Sigma^2+2QP}$, as an inner horizon, and the area of the two-sphere shrinks to zero at $r_{S}$. Moreover, it is interesting that when the dilaton charge is zero, one obtains a non-extremal black hole, with horizon at $r_{+}= +\sqrt{2QP}$. To see this one can argue that massless solution seems physically acceptable, even with a complex dilaton field at infinity.

Noting Eq. (\ref{aa}), to describe the full massless nonextremal solution, one must choose the negative sign in Eq. (\ref{a}) with constants $d_{1}= -d_{0}= -\Sigma$. By so doing, we find the expression (\ref{aa}) leads to \cite{Goulart2}
\begin{eqnarray}\label{metric}
ds^2 &=& -\left(\frac{r^2}{r^2+2PQ}\right)\mathrm{d}t^2
+\left(\frac{r^2+2PQ}{r^2+\Sigma^2+2PQ}\right)dr^2 \nonumber\\
&& +(r^2+2QP)(d\theta^2+\sin^2\theta d\varphi^2).
\end{eqnarray}
The above metric represents the charged wormhole in EMD theory,
which can be obtained from the massless nonextremal dyonic solution. Also, once we pick a value 
for $\Sigma =0$, we recover the solution of the Einstein-Rosen bridge with throat radius 
$\mathcal{R_{\text{thro}}}=\sqrt{2PQ}$.
\begin{figure}
  \includegraphics[scale=0.65]{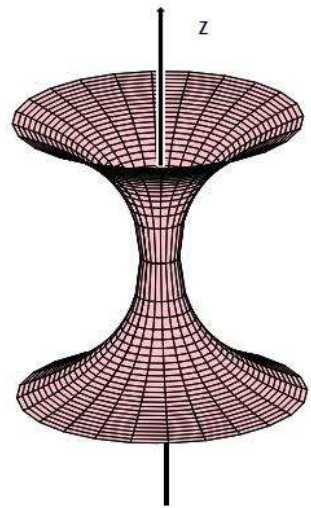}\\
  \caption{The embedding diagram of a two-dimensional section along the equatorial plane ($t$ = const, $\theta$ = $\pi$/2) of a charged wormhole solution in EMD theory. To visualize this we plot $z$ vs. $r^{\star}$ sweep through a 2$\pi$ rotation around the $z$-axis.}\label{f1}
\end{figure}

This line element indicates that there is no singularity in this
spacetime. To understand the geometry of the spacetime
intuitively, we draw an embedding diagram as follows. Introducing a new radial coordinate as
\begin{eqnarray}
r^{\star}\equiv \sqrt{r^2+2PQ}.
\end{eqnarray}
The respective line element (\ref{metric}), for a fixed moment of $t$ = const, and  $\theta =\pi/2$, which 
yield
\begin{eqnarray}\label{cw}
ds_{\text{CW}}^2= \frac{d{r^{\star}}^2}{\left(1-\frac{2PQ}{{r^{\star}}^2}\right)
\left(1+\frac{\Sigma^2}{{r^{\star}}^2}\right)} +{r^{\star}}^2 d\varphi^2.
\end{eqnarray} 
To describe the whole scenario, one embeds this metric into three-dimensional Euclidean
space described by cylindrical coordinates ($r$, $\varphi$, $z$) as
\begin{eqnarray}\label{ew}
ds_{\text{EW}}^2= dz^2 +d{r^{\star}}^2 +{r^{\star}}^2 d\varphi^2 .
\end{eqnarray}
Assuming the line elements (\ref{cw}) and (\ref{ew}) are equal, i.e. 
$ds_{\text{CW}}^2= ds_{\text{EW}}^2$, and then one can 
obtain a simple relation for a specific value when $\Sigma =0$, in the following form
\begin{eqnarray}
z= \pm \sqrt{2PQ}~ \text{arccosh} \frac{r^{\star}}{\sqrt{2PQ}} .
\end{eqnarray}
 This can be visualized in Fig.~\ref{f1} through a $2\pi$ rotation around the 
$z$-axis. Moreover, we can see from the Fig.~\ref{f1} that two separated spacetimes are connected by the
throat. Far from the throat, the spacetime is asymptotically flat, i.e., $dz/d{r^{\star}}\rightarrow 0$
as $r^{\star}\rightarrow \infty$.

It is possible to construct the massless black holes for the EMD theory whose observables 
are real \cite{Goulart1}. Furthermore, using the same massless solutions one can construct Einstein-Rosen 
bridges which satisfy the null energy condition. Motivated by this theory, in the next section, 
we study motion of the photons in the gravitational field of charged wormholes within the context of 
the EMD theory.

\begin{figure*}
\includegraphics[scale=0.65]{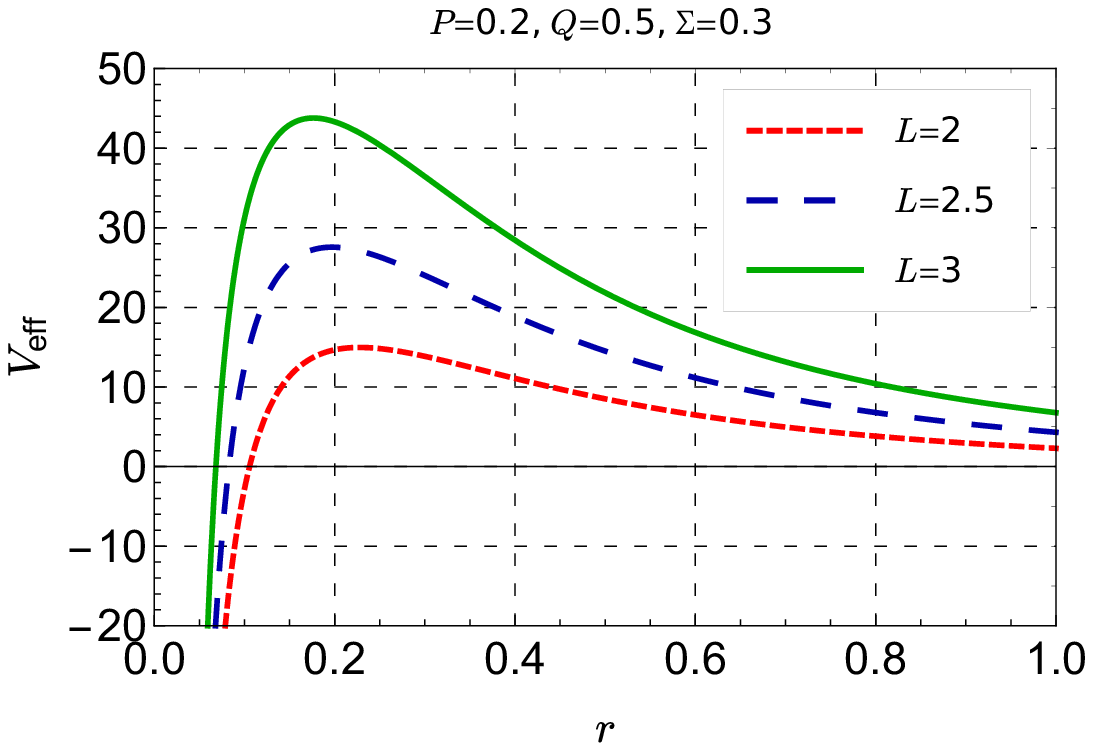}
\includegraphics[scale=0.65]{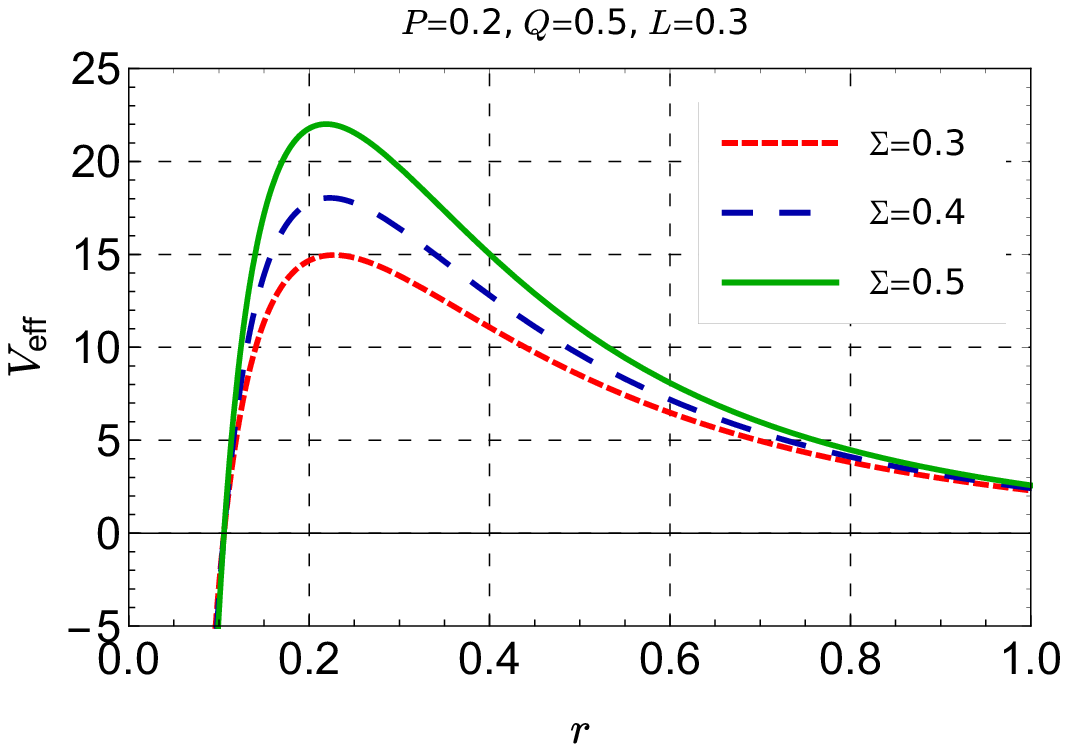}
\caption{\label{eff} Plots showing the nature of effective potential for charged wormhole in EMD gravity by 
varying the angular momentum $L$ and dilaton charge $\Sigma$. Here the unit of length is arbitrary.}
\end{figure*}

\section{Photon motion around charged wormholes in EMD theory}
\label{geodesics}
\begin{figure}
\includegraphics[scale=0.35]{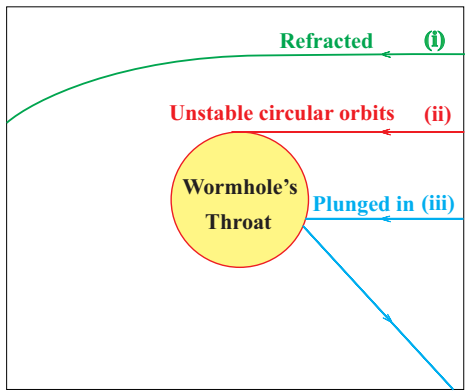}
\caption{\label{traject} Schematic representation of the photon trajectories around the wormholes.}
\end{figure}

In this section, our aim is to discuss the motion of photons around charged wormholes that can be described by evaluating 
the geodesic equations. To determine the photon geodesics, we consider the conserved quantities associated with the 
spacetime metric of charged wormholes in EMD theory. It is clear that the spacetime metric (\ref{metric}) is invariant 
under time translation and spatial rotation which leads to the conservation of energy and conservation of angular momentum, 
respectively. Consequently, it is characterized by two constants of motion, i.e., energy $E=-p_t$ and angular momentum 
$L=p_{\varphi}$. Now we can derive the geodesic equations by using these conserved quantities
\begin{eqnarray}\label{u}
\frac{dt}{d\sigma} &=& \frac{E(r^2+2PQ)}{r^2}, \nonumber \\
\frac{d\varphi}{d\sigma} &=&  \frac{L\csc^2 \theta}{r^2 + 2PQ},
\end{eqnarray}
where $\sigma$ indicates the affine parameter along the geodesic. Since  test particle geodesics  
around the wormholes satisfy the Hamilton-Jacobi equation
\begin{equation}\label{hje}
\frac{\partial S}{\partial \sigma} = -\frac{1}{2} g^{\mu \nu} 
\frac{\partial S}{\partial x^{\mu}} \frac{\partial S}{\partial x^{\nu}},
\end{equation}
and for photons ($m_0=0$),  Eq.~(\ref{hje}) has a solution of the following form
\begin{equation}\label{s}
S = -Et +L \varphi +S_r(r) +S_{\theta}(\theta),
\end{equation}
where $S_r$ and $S_{\theta}$ are the functions of $r$ and $\theta$, respectively. On substituting (\ref{s}) into (\ref{hje}) as well as contravariant component of metric, i.e., $g^{\mu \nu}$, 
and separating the terms of variables $r$ and $\theta$ equal to the Carter constant 
($\pm \mathcal{K}$) \cite{Carter:1968rr}, we obtain
\begin{eqnarray}
\frac{r\sqrt{r^2+2PQ}}{\sqrt{r^2+\Sigma^2 +2PQ}}\frac{dr}{d\sigma} &=& 
\pm \sqrt{\mathcal{R}}, \nonumber \\
(r^2+2PQ)\frac{d\theta}{d\sigma} &=&  \pm \sqrt{\Theta},
\end{eqnarray}
where $\mathcal{R}$ and $\Theta$ have the following forms
\begin{eqnarray}
\label{quant}
\mathcal{R} &=& E^2(r^2 + 2PQ) -\frac{r^2}{(r^2 + 2PQ)}(\mathcal{K}+L^2), 
\nonumber \\
\Theta &=& \mathcal{K} -L^2 \cot^2 \theta.
\end{eqnarray}
Having the geodesic equations, we turn our attention towards the radial motion of photons around 
the throat of charged wormholes, which can be demonstrated by the calculation of the effective 
potential
\begin{eqnarray}
\left(\frac{dr}{d\sigma}\right)^2 + V_{eff}=0,
\end{eqnarray}
with
\begin{eqnarray}\label{V}
V_{eff} = -\frac{r^2+\Sigma^2 +2PQ}{r^2}\left[E^2 -\frac{r^2(\mathcal{K} +L^2)}{(r^2 + 2PQ)^2}\right].
\end{eqnarray}
It is clear from the expression of the effective potential that it has dependency on the charges $P$, $Q$, and 
$\Sigma$ as well as on angular momentum $L$. We can visualize the behavior of the effective potential from 
Fig.~\ref{eff} for different values of angular momentum $L$ and dilaton charge $\Sigma$. When there is an increase 
in the value of angular momentum $L$ and dilaton charge $\Sigma$, it turns out that there is an increase in the peak 
of effective potential where the unstable circular orbits form. There occurs three kinds of photon trajectories 
around the throat of wormholes: (i) refracted by the gravity, or (ii) form circular orbits, or (iii) traversed to 
the other region (cf. Fig.~\ref{traject}). 

It is convenient to reduce the number of parameters by defining the impact parameters such that $\xi=L/E$ and 
$\eta=\mathcal{K}/E^2$. Photon geodesics can be expressed in terms of these impact parameters ($\xi$, $\eta$). We can 
rewrite $\mathcal{R}$, in terms of impact parameters as follows
\begin{eqnarray}
\label{R}
\mathcal{R} &=& -E^2 \left[r^2 + 2PQ 
-\frac{r^2(\mathcal{\eta} +\xi^2)}{(r^2 + 2PQ)}\right].
\end{eqnarray}
The study of effective potential reveals the existence of unstable circular orbits of constant 
radius around the throat of wormholes. These orbits are very important from the point of view of 
optical observations.

\section{Shadow of charged wormholes}
\label{wshadow}
In this section, we investigate the shadow cast by the charged wormholes in EMD theory. As discussed earlier the 
wormhole is a tunnel-like structure connecting different regions of spacetime and it does not contain any horizon 
or singularity within it. Moreover, the wormholes have gravitational tidal forces so weak that can be assumed to be 
bearable by a human. The throat of the wormholes plays a crucial role, as we showed earlier there exist unstable photon 
orbits revolving around it infinite times before reaching the observer.  

We consider the situation where one of the spacetime regions that is connected by the wormholes is illuminated by 
a light source and a distant observer is also placed in this region. One should note that there is no light source in 
another spacetime region therefore no photons traversing from this region. Since a distant observer can be able to observe 
the scattered photons. However, the photons which are plunging, consequently traversing through the wormholes cannot be 
observed by the observer. The absence of the traversed photons forms a dark spot on the luminous background known as shadow 
of wormhole. The unstable circular photon orbits give the boundary of the shadow. Equations that determine the unstable 
circular orbits can be expressed as 
\begin{equation}\label{condition}
\mathcal{R} = 0, \quad  \mathcal{R}'=0,
\end{equation}
where prime ($'$) represents derivative with respect to $r$. Substituting (\ref{R}) into 
(\ref{condition}), one can easily derive
\begin{eqnarray}\label{imptp}
\eta  &=& 8PQ - \xi ^2, \quad r = \sqrt{2PQ}.
\end{eqnarray}
Note that the impact parameter $\eta$ has a dependency on the magnetic charge $P$ as well as on the electric 
charge $Q$, but it is independent with the dilaton charge $\Sigma$. 

\begin{figure*}
\includegraphics[scale=0.535]{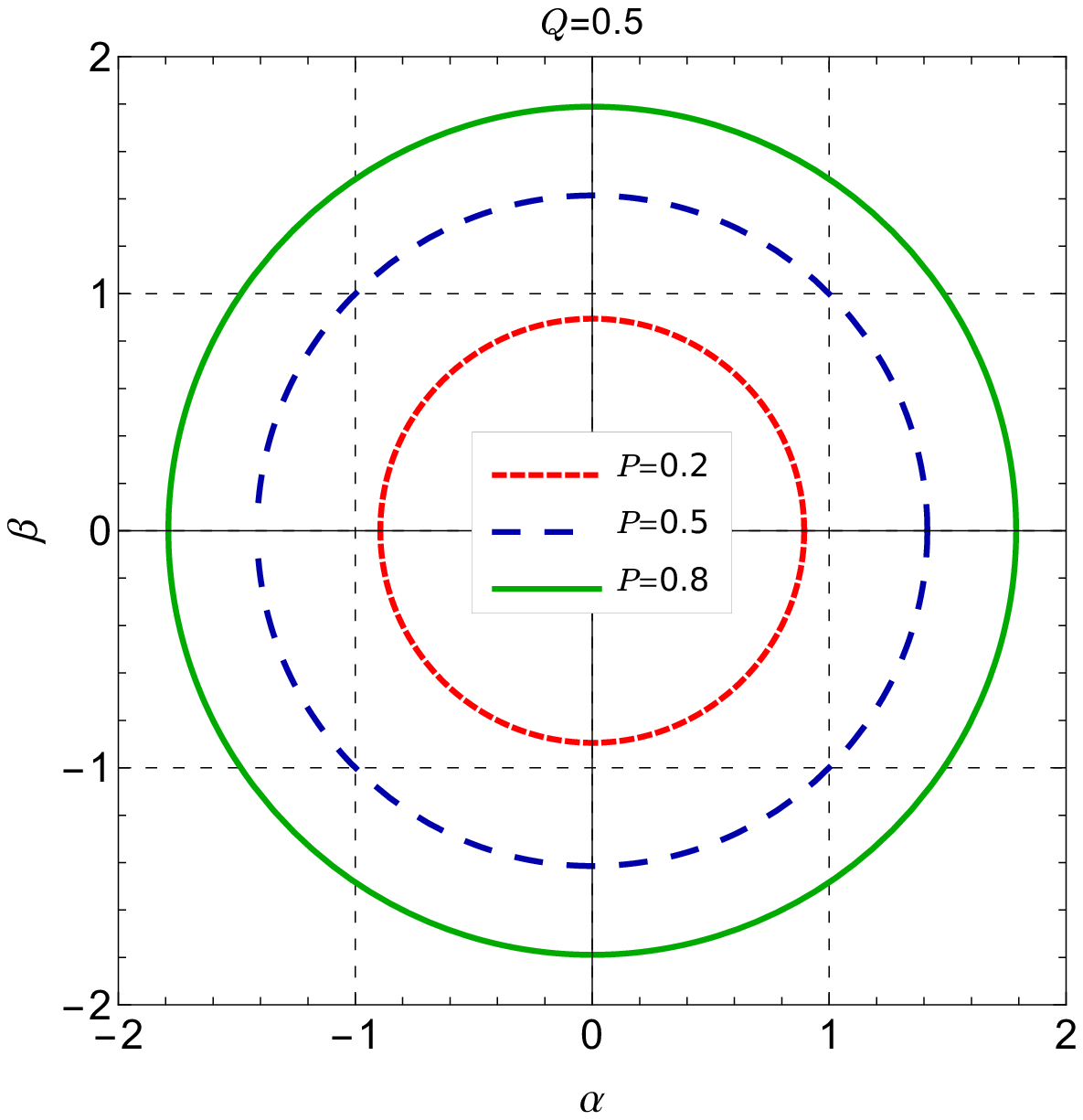}
\includegraphics[scale=0.555]{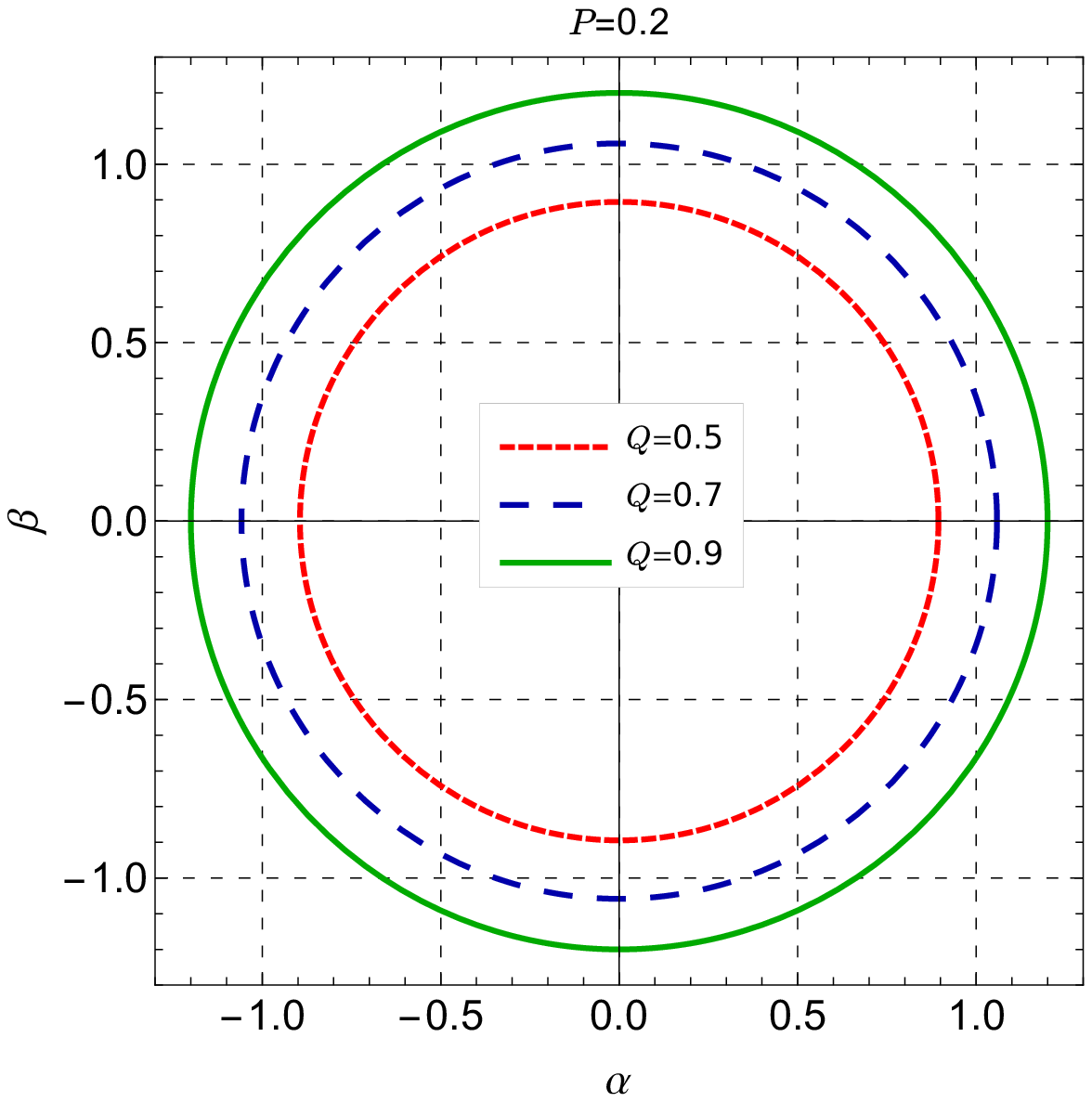}
\caption{\label{shadow} Plots showing the shapes of the shadow of charged wormhole in EMD theory by varying the 
charges with $\theta_0=\pi/2$. Here the unit of length is arbitrary.}
\end{figure*}

Equation~(\ref{imptp}) determines the boundary of shadow and a distant observer can observe its 
projection in `his sky'. Therefore one needs to define the celestial coordinates ($\alpha, \beta$) in 
the observer's sky and relate them with impact parameters ($\xi$, $\eta$). These celestial 
coordinates can be defined \cite{Bardeen:1973gb,Chandrasekhar:1992} as  
\begin{eqnarray}
\label{celstf}
\alpha &=& \lim_{r_0 \rightarrow \infty}\left(r_0^2 \sin \theta_0 \frac{d\varphi}{dr}\right), \nonumber\\
\beta &=& \lim_{r_0 \rightarrow \infty} r_0^2 \frac{d\theta}{dr},
\end{eqnarray}
where $r_0$ is the distance from the wormholes to the observer and $\theta_0$ is the angular 
coordinate of the observer or one can say the inclination angle. On substituting the expressions of 
four-velocities into (\ref{celstf}), and after some straight forward calculations, we obtain the 
following forms of celestial coordinates
\begin{eqnarray}
\label{celst}
\alpha &=& -\xi \csc \theta_0, \quad \beta = \pm \sqrt{\eta- \xi^2 \cot^2 \theta_0}.
\end{eqnarray}
Having the expressions of celestial coordinates and impact parameters, we are in a position to construct the 
shadow of charged wormholes. In order to obtains the shapes of shadow, we plot $\alpha$ vs $\beta$ that gives the 
boundary of shadow in the observer's sky. The shapes of the shadow for the charged wormholes can be visualized from 
Fig.~\ref{shadow}. We have shown the behavior of shadow by varying the magnetic ($P$) and electric ($Q$) charge in 
the equatorial plane $\theta_0 =\pi/2$. We find that the shape of the shadow is a perfect circle and it is affected 
by the presence of charges (cf. Fig.~\ref{shadow}). As a consequence we can say that the shadow of charged 
wormholes increases with both of the charges $P$ and $Q$.

\section{Conclusion}
\label{concl}
Since last few decades the study of shadow cast by the compact objects especially for black holes and wormholes 
has been discussed prominently. As the subject has astrophysical significance because strong evidences suggest the 
presence of supermassive black holes at the hearts of giant galaxies. The study is very important from the point 
of view of discussing the nature of these astrophysical objects. As we know that wormholes are the most exotic 
objects predicted by general relativity which act as tunnel linking between two spacetimes as a two-way interface, 
in the same (or not) Universe. In this paper, we have constructed the shadow of charged wormholes in EMD theory. 
The spacetime metric of charged wormholes has three different types of charges, namely, magnetic charge $P$, 
electric charge $Q$, and dilaton charge $\Sigma$. We have calculated the photon geodesics and explained the 
possible trajectories of them around the charged wormholes in EMD theory. Furthermore, we have discussed the 
behavior of the effective potential with angular momentum $L$ and dilaton charge $\Sigma$. We have evaluated the 
impact parameters and celestial coordinates to image the boundary of shadow for the charged wormholes. We found 
that the shapes of charged wormholes shadow are the perfect circles and they are affected by the charges $P$ and 
$Q$. Eventually, to visualize the effect of these charges on the shape of wormholes shadow, we have constructed 
the graphical illustration for different cases. As a consequence we found that the radius of the shadow increases 
when there is a small increase in the magnetic charge $P$ as well as in the electric charge $Q$. Apart from it, 
we observed that the radius of the shadow has not any dependency on the dilaton charge $\Sigma$. It turns out that 
we cannot determine the dilaton charge by the shadow observations. We expect that in upcoming years the direct 
images of wormholes will be test by the Event Horizon Telescope.

\begin{acknowledgments}
M.A. and A.B. would like to thank University of KwaZulu-Natal and the National Research 
Foundation for financial support. SDM acknowledges that this work is based on research supported by 
the South African Research Chair Initiative of the Department of Science and Technology and the National 
Research Foundation. We would like to thank the referee for useful comments and suggestions.
\end{acknowledgments}

\end{document}